\documentclass{article}
\usepackage{amsmath,graphicx,mlspconf}
\usepackage{multirow}
\usepackage{booktabs}
\usepackage{amsmath}

\title{T-CLAP: Temporal-Enhanced Contrastive Language-Audio Pretraining}
\name{
\parbox{\linewidth}{\centering
   Yi Yuan$^{1}$,%
   Zhuo Chen$^{2}$,%
   Xubo Liu$^{1}$,%
   Haohe Liu$^{1}$,%
   Xuenan Xu$^{3}$,%
   Dongya Jia$^{2}$,%
   Yuanzhe Chen$^{2}$ %
   \\ Mark D. Plumbley$^{1}$,%
   Wenwu Wang$^{1}$%
   }
}
\address{%
   $^{1}$ University of Surrey, United Kingdom %
   $^{2}$ ByteDance, China \\ %
   $^{3}$ Shanghai Jiao Tong University, China %
}

\begin{document}

\maketitle

\begin{abstract}

\noindent Contrastive language-audio pretraining~(CLAP) has been developed to align the representations of audio and language, achieving remarkable performance in retrieval and classification tasks. However, current CLAP struggles to capture temporal information within audio and text features, presenting substantial limitations for tasks such as audio retrieval and generation. To address this gap, we introduce T-CLAP, a temporal-enhanced CLAP model. We use Large Language Models~(LLMs) and mixed-up strategies to generate temporal-contrastive captions for audio clips from extensive audio-text datasets. Subsequently, a new temporal-focused contrastive loss is designed to fine-tune the CLAP model by incorporating these synthetic data. We conduct comprehensive experiments and analysis in multiple downstream tasks. T-CLAP shows improved capability in capturing the temporal relationship of sound events and outperforms state-of-the-art models by a significant margin. Our code and models will be released soon.

\end{abstract}

\begin{keywords}
    Contrastive language-audio pretraining, audio retrieval, zero-shot classification, multi-modal learning
\end{keywords}

\section{Introduction}
\label{sec:introduction}
Contrastive pretraining has received wide attention in the field of multi-modal learning, such as contrastive language-image pretraining~(CLIP)~\cite{clip}. By training the models with a large number of paired samples, such as text-image pairs, these models align the latent representations across the modalities, showing excellent performance in various language-related downstream tasks, including classification~\cite{class1,class2}, retrieval~\cite{retrieval1,retrieval2}, captioning~\cite{caption1,caption2}, and generation~\cite{generate1,generate2,generate3}. Similar to CLIP, the recently proposed CLAP model~\cite{clap} applies an audio encoder and a text encoder to learn the correspondence between text description and audio signal in a shared latent space, setting state-of-the-art performance for audio retrieval tasks. Recently, the ability of CLAP to capture rich auditory features through both audio and text inputs has become a key component in text-to-audio models~\cite{audioldm,reaudioldm}.


However, experiments indicate that current contrastive models exhibit weak modelling capabilities of temporal information~\cite{pre_exp}. For example, when processing an audio clip that features a sequence of events, such as ``\textit{A dog barking followed by a man speaking}'', CLAP often generates embeddings that do not distinguish between this sequence and the sequence that occurs in reverse order, i.e. ``\textit{Man speaks followed by dog barks}''. We hypothesize that the existing deficiency is caused by several factors. Firstly, previous research~\cite{pre_search} indicates that transformer-based encoders, such as those applied in CLAP models, are not sensitive to temporal information.
Secondly, the existing datasets are deficient in samples that are similar in auditory features but different in sequential order~\cite{audiocaps_statis}. This scarcity affects the ability of CLAP to recognize the temporal features effectively. In addition, part of the training data used to develop the CLAP models are captions augmented from keywords~\cite{clap}, which lack sequential features that could potentially drive the model towards exploiting temporal information. 
These issues limit the model's ability to parse and represent the correct order of audio events in multi-event scenarios. As a result, its performance in downstream tasks that involve temporal ordering may be poor. For example, audio retrieval systems built on the CLAP models may fail to correctly retrieve an audio clip in which the order of the audio events is specified by the input text. Similarly, audio generation models built on CLAP models may generate audio clips with events in the wrong order. These examples demonstrate the strong demand for improved CLAP models that can capture temporal information (e.g., temporal order of audio events) within the audio and text data.

In this work, we propose a straightforward but novel approach, T-CLAP, to enhance the temporal understanding of CLAP models. In detail, several approaches are designed to generate samples with temporal contrastive information~(i.e. audio events with different sequential order). We first create samples by mixing up labels and audio. Secondly, we take original texts as positive captions and leverage Large Language Models (LLMs) to generate negative captions for audio clips in the large-scale audio-text paired datasets. Then a temporal-focused loss is proposed to fine-tune the CLAP model with these additional text data. We further present a new task called T-Classify to assess the temporal information captured within CLAP embeddings. T-Classify is designed to examine the ability of the CLAP models to identify and retrieve audio and text data with accurate temporal information.


Our experimental results show that T-CLAP outperforms baseline CLAP models and achieves the state-of-the-art in T-Classify, audio-text retrieval and zero-shot classification tasks. To further evaluate the representation of T-CLAP, we conduct experiment generation tasks (e,g. text-to-audio generation). We use different CLAP models as the text encoder for AudioLDM~\cite{audioldm}, a state-of-the-art latent diffusion-based generation model. Both objective and subjective evaluation results demonstrate that T-CLAP improves the performance of AudioLDM in generating sound events aligned with the temporal relationships described in the text captions. 


The paper is organized as follows. Section~\ref{sec:tcalp} introduces the T-CLAP model, including the structure of the CLAP model, the proposed data processing pipelines and the loss functions. Section~\ref{sec:experiment} presents the experimental settings and provides the evaluation metrics of the model. Section~\ref{sec:result} discusses the results and the conclusion is drawn in Section~\ref{sec:conclusion}. 

\section{Temporal-Enhanced CLAP}
\label{sec:tcalp}
\subsection{Problem Formulation}
Let $ G = \left\{ (a_i, t_i) \right\}_{i=1}^N $ be a dataset with $N$ audio-caption samples, where the audio clip $a_i$ and its associated audio caption $t_i$ form an audio-text pair indexed by $i$. 
The CLAP utilizes an audio encoder $f_{\mathrm{audio}}(\cdot)$ and a text encoder $f_{\mathrm{text}}(\cdot)$ to extract the feature of audio clips and their captions, and then projects them into a shared dimension $D$, i.e., $f_{\mathrm{text}}(a_i)\in {R}^{D}$ and $ f_{\mathrm{audio}}(t_i)\in {R}^{D}$. Using cosine similarity, the similarity of each audio and text embedding can be measured as: 
\begin{align}
    s_{ij} = \frac{f_{\mathrm{text}}(a_i) \cdot f_{\mathrm{audio}}(t_j)}{\begin{Vmatrix}f_{\mathrm{text}}(a_i)\end{Vmatrix}_2 \cdot \begin{Vmatrix}f_{\mathrm{audio}}(t_j)\end{Vmatrix}_2}
\end{align}

\noindent where the CLAP is trained to increase the similarity score of each audio with its matched text $s_{ii}$, while decreasing similarity with other text $s_{ij}~(i \neq j)$.As discussed in Section~\ref{sec:introduction}, original CLAP models suffer from data limitations and encoder insensitivity in temporal features, resulting in insufficient performance in audio event order understanding and generation.

\subsection{Model}
Our goal is to train an enhanced CLAP that performs better on capturing temporal features, i.e., presenting the text embedding with correct sequential information.
We leverage transfer learning and utilize a pretrained CLAP~\cite{clap} to learn the feature with temporal information. In detail, we employ HTSAT~\cite{htsat} as the audio encoder and RoBERTa~\cite{roberta} as the text encoder. The multi-layer perceptron (MLP) layers and activation functions are then applied to project audio and text embedding into the same dimension for contrastive learning. 

\subsection{Training Set}
\label{sec:trainingset}
To enhance the model in capturing temporal information, we introduce two novel approaches to generate temporally contrastive captions (e.g., captions describing audio events in a wrong order). We first generate audio clips with different audio events by mixing and concatenating audio clips and creating their corresponding text descriptions with matching temporal ordering phrases. 
\begin{figure}[htbp]
  \centering
  \includegraphics[width=\linewidth]{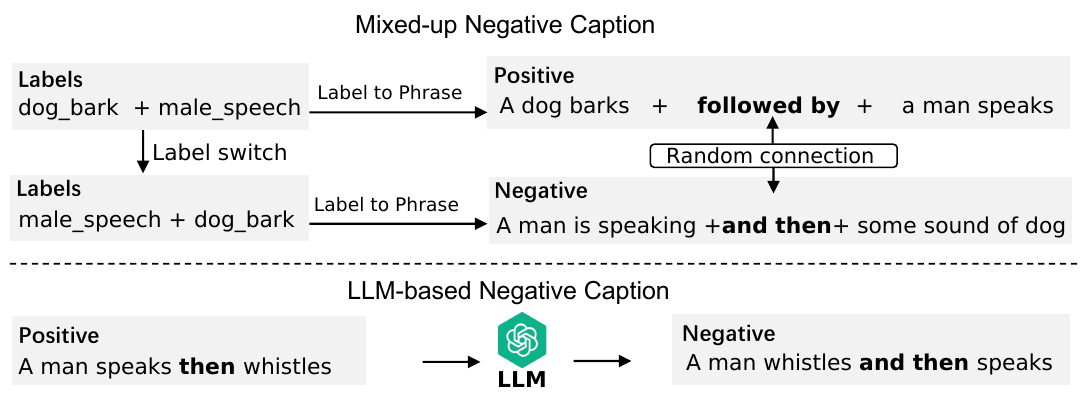}
  \caption{Pipelines of generating the negative captions.}
  \label{fig:figure1}
\end{figure}

\noindent As shown in Figure~\ref{fig:figure1}, labels of each audio are converted into short captions connected with phrases, such as {\textit{``\textit{followed by}'', ``\textit{and then}''}}, to represent the temporal ordering of audio events. A negative caption is then generated by switching the order of the description for each audio event within each sentence. 

\begin{figure}[htbp]
  \centering
  \includegraphics[width=0.9\linewidth]{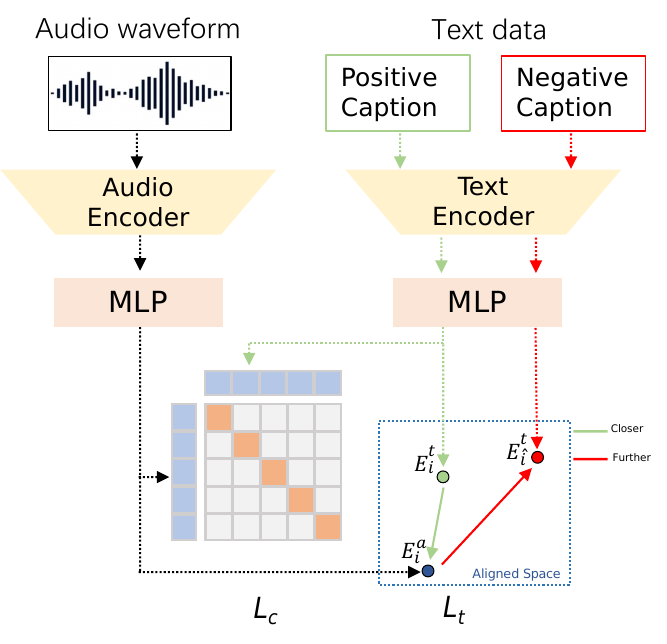}
  \vspace{-10pt}
  \caption{Pipeline for training T-CLAP, with original contrastive loss $L_{c}$ on the left and proposed temporal-focused loss $L_{t}$ on the right.}
  \label{fig:figure2}
  \vspace{-15pt}
\end{figure}

Secondly, we apply LLMs, e.g. ChatGPT, to generate negative captions of existing large-scale audio-caption datasets. The LLMs are guided to identify the descriptions for audio events in each sentence and generate the text with the same audio events in a different order. We have observed that this approach provides more diversity of temporal-contrastive samples than artificially generated data~(e.g., mixed-up audio and caption), which cannot cover various conditions and sentence combinations.

Nevertheless, we found that audio captions generated by LLMs tend to exhibit specific characteristics. For example, a substantial number of negative captions consist of audio events connected with words like ``first'' or structures like `` , '' to represent temporal information, leading the model to overfit on these specific conditions. Hence, we combine both mixed-up datasets and LLM-generated datasets for training, where mixed-up datasets provide straightforward contrastive samples and LLM-generated datasets provide comprehensive and diverse samples. Statistic details of the datasets are discussed in Section~\ref{sec:experiment}.


\begin{table*}[htbp]\small
\caption{The retrieval performance between different systems, CLAP$_{m}$ and CLAP$_{l}$ are models trained by Microsoft~\cite{oldclap,CLAP-micro} and LAION~\cite{clap}, using different structures and datasets respectively~\cite{clap}. For the training set, ``AC'' and ``CL'' are short for AudioCaps and Clotho datasets, ``WT5K'', ``LA'', ``ES'' and ``AU'' refer to Wavtext5K, LAION-Audio-630k,  ESC-mixed-up and Auto-ACD respectively. T-CLAP is marked with $\dagger$ as a fine-tuned model  from CLAP$_{l}$ }
\centering
\resizebox{1.01\textwidth}{!}{%
\begin{tabular}{cccccccc|cccccc}
\toprule
\multirow{4}{*}{\textbf{Model}} & \multirow{4}{*}{\textbf{Training Set}} & \multicolumn{6}{c}{\textbf{AudioCaps}} & \multicolumn{6}{c}{\textbf{Clotho}}  \\
\cmidrule(lr){3-14}
& & \multicolumn{3}{c}{\textbf{Text-to-Audio}} & \multicolumn{3}{c}{\textbf{Audio-to-Text}} & \multicolumn{3}{c}{\textbf{Text-to-Audio}} & \multicolumn{3}{c}{\textbf{Audio-to-Text}} \\ 
\cmidrule(lr){3-14}
& & \textbf{R@1} & \textbf{R@5} & \textbf{R@10} & \textbf{R@1} & \textbf{R@5} & \multicolumn{1}{c}{\textbf{R@10}} & \multicolumn{1}{c}{\textbf{R@1}} & \textbf{R@5} & \textbf{R@10} & \textbf{R@1} & \textbf{R@5} & \textbf{R@10} \\ 
\midrule

MMT~\cite{MMt} & AC or CL & $36.1$  & $72.0$ & $84.5$ & $39.6$ & $76.8$ & $86.7$ & $6.7$  & $21.6$ & $33.2$ & $7.0$ & $22.7$ & $34.6$ \\

ML-ACT~\cite{mlase} & AC or CL & $33.9$  & $69.7$ & $82.6$ & $39.4$ & $72.0$ & $83.9$ & $14.4$  & $36.6$ & $49.9$ & $16.2$ & $37.6$ & $50.2$ \\

TAP~\cite{TAP} & AC or CL & $36.1$  & $72.0$ & $85.2$ & $41.3$ & $75.5$ & $86.1$ & $16.2$  & $39.2$ & $50.8$ & $17.6$ & $39.6$ & $51.4$ \\

CLAP~\cite{oldclap} & AC+CL+WT5K & $34.6$  & $70.2$ & $82.0$ & $41.9$ & $73.1$ & $84.6$ & $16.7$  & $41.1$ & $\textbf{54.1}$ & $20.0$ & $44.9$ & $58.7$ \\

CLAP$_{m}$~\cite{CLAP-micro} & 4.6M-Audio  & $33.5$  & $70.4$ & $80.2$ & $47.8$ & $80.2$ & $90.7$ & $16.2$  & $39.6$ & $51.4$ & $\textbf{23.6}$ & $46.7$ & $\textbf{60.3}$ \\


\midrule
CLAP$_{l}$~\cite{clap} & AC+CL+LA & $34.2$  & $71.1$ & $84.1$ & $43.1$ & $79.5$ & $90.1$ & $15.3$  & $38.4$ & $51.2$ & $20.8$ & $\textbf{51.2}$ & $60.0$ \\

T-CLAP$^\dagger$ & AC+CL+ES+AU &  $\textbf{39.7}$&  $\textbf{74.6}$ & $\textbf{86.9}$ & $\textbf{49.8}$ & $\textbf{82.5}$ & $\textbf{91.9}$ & $\textbf{17.3}$&  $\textbf{39.9}$ & $53.6$ & $21.8$ & $44.9$ & $57.4$ \\

\bottomrule
\end{tabular}
}
\label{tab:retrieval}
\end{table*}

\subsection{Loss Functions}
\label{loss}
The key idea is to train the model to embed audio and text features with temporal information. This means that an audio caption describing the correct order of audio events results in a higher similarity score than a caption describing audio events in a wrong (or different) order. Inspired by \cite{countclip}, we introduce a temporal-focused loss that guides the model to minimize the distance between target audio and positive caption embedding within the aligned latent space, while pushing it away from the embedding presented in the negative order. To effectively enhance our model and maintain the retrieval capabilities, we use a combination of two loss functions for fine-tuning CLAP: 

\begin{align}
    L_{\mathrm{train}} = L_{c} + \lambda_{l} L_{t}
\end{align}

\noindent where $ L_{c}$ is the contrastive loss that maintains the capability of the CLAP model, while $L_{t}$ is the newly introduced temporal-focused loss for guiding the model with temporal information, and $\lambda_{l}$ is a weight to balance the two losses. As shown in Figure~\ref{fig:figure2}, for each audio-text sample $i$, the model first computes the aligned-presentation of audio embedding $E^{a}_{i}$, the positive text embedding $E^{t}_{i}$ and negative text embedding $E^{t}_{\hat{i}}$. The term $L_{t}$ then works as a single-contrastive-loss to reinforce the similarity between audio and positive text, while minimizing it with negative text, shown as: 
\begin{align}
    L_{t} = - \sum_{i=1}^{N}\log \frac{\exp(E^{a}_{i} \cdot E^{t}_{i})}{\exp(E^{a}_{i} \cdot E^{t}_{i}) + \exp(E^{a}_{i} \cdot E^{t}_{\hat{i}})}
\end{align}

\section{Experiments}
\label{sec:experiment}

\subsection{Dataset}
\textbf{Training set.} In this paper, we first utilize ESC-50~\cite{esc-50}, which comprises $50$ distinct categories of sounds with each audio sample lasting $5$ seconds, as the supplementary database for mix-up negative captions~(ESC-mixed-up). Applying the dataset generation pipeline introduced in Section~\ref{sec:trainingset}, the ESC-mixed-up dataset consists of $50,000$ 10-second audio clips. Each audio clip is composed of distinct audio events and matched with one positive and one negative caption. In addition, we employ LLMs to generate the negative texts corresponding to captions in AudioCaps~\cite{audiocaps} and Clotho~\cite{clotho}. Both AudioCaps and Clotho are audio datasets with human-annotated captions. Specifically, AudioCaps includes about $50,000$ audio samples, each lasting 10 seconds, and Clotho has about $5,000$ audio clips of varying duration~(ranging from $15$ to $30$ seconds). In addition, Auto-ACD~\cite{autoacd} is applied as the large training set to maintain the retrieval capability of T-CLAP. Auto-ACD contains $1.9$M audio samples with paired captions generated through LLMs on both visual and audio information. Together we collected $2.04$M audio-caption pairs for training.

\noindent \textbf{Testing set.} We evaluate the performance of the proposed method on several downstream tasks, including retrieval, zero-shot classification, and text-to-audio generation. For the retrieval tasks, we follow the baseline models and evaluate AudioCaps and Clotho.
For temporal-focused retrieval~(T-Classify), we apply AudioCaps, Clotho and ESC-mixed-up datasets with temporal-negative samples.
For zero-shot classification tasks, we use ESC-50, Urbansound8k~\cite{ub8k} and VGGSound~\cite{vggsound} for evaluations. Urbansound8K has $8,000$ samples and is labelled into ten different groups, while VGGSound is a larger dataset with more than $300$ classes.

\subsection{Implementation}
We fine-tune the CLAP~\cite{clap} with our proposed loss function $L_{\mathrm{train}}$ introduced in Section~\ref{loss}.
Specifically, the temporal loss $L_{t}$ is only applied on a combination of AudioCaps, Clotho, and ESC-mixed-up datasets, where each sample is combined with a negative caption. We take the Auto-ACD with only positive captions as the primary database for the contrastive loss $L_{c}$ and apply a ratio of $1:4$ to add the other three datasets within each batch. The weight between two loss functions $\lambda_{l}$ is set to $0.5$. All the models are trained over $30$K steps with a batch size of $512$. The learning rate is set to $1\times10^{-4}$ with a linear warm-up for the first $10,000$ steps. 

\subsection{Evaluation}
Four different tasks are applied to evaluate the performance of T-CLAP on different tasks. First, we follow the baseline models on retrieval tasks, including text-to-audio (T2A) and audio-to-text (A2T) retrieval tasks. 

In addition, we evaluate the performance of the T-CLAP model on zero-shot audio classification across three datasets, including ESC-50~\cite{esc-50}, Urbansound8K~\cite{ub8k}, and VGGSound~\cite{vggsound}. For ESC-50, we follow the original test-split defined within the CLAP~\cite{clap} experiments, while we examine the performance on the entire Urbansound8K dataset and the official testing set of VGGSound, respectively. Similar to the baseline approach, we evaluate the performance as a T2A retrieval task with a text prompt formatted as `` \textit{A sound of \textbf{Label}} '' according to the \textit{\textbf{Label}} of each audio sample. Notably, we have excluded all the testing samples from the ESC-50 dataset in our mixed-up dataset and VGGSound audio samples in Auto-ACD to prevent any overlap and premature access of data during the training stages. 

We also introduce a new task, T-Classify, to measure the ability of temporal feature retrieval.
For each test sample $i$ with negative captions~(AudioCaps and Clotho testing set), T-Classify evaluates the performance on T2A by calculating the percentage of cases where the model can successfully provide higher cosine similarity on the positive captions and lower on negative captions~($s_{ii} > s_{i\hat{i}}$). Same as for A2T tasks, for each test sample $j$ with negative audio clips~(ESC-mixed-up testing set), T-Classify measures the percentage that the model calculates a higher score on positive audios than negative audios~($s_{jj} > s_{\hat{j}j}$). For both two tasks, we calculate the percentage of correct judgments by aggregating the results across the testing sets. 

Lastly, we evaluate the effectiveness of temporal-enhanced embedding in text-to-audio generation tasks. Specifically, we train various 
AudioLDM~\cite{audioldm} using different CLAP models as the input encoders.
In addition to objective metrics on previous experiments for audio generation, we conduct human evaluations to assess if the AudioLDM, trained with the enhanced CLAP, can generate outputs whose temporal features align with the given text conditions.

\section{Results}
\label{sec:result}
All the results are evaluated based on an average of each experiment trained three times with random seeds. 

\noindent \textbf{Text and audio retrieval.} Table~\ref{tab:retrieval} presents the performance comparison of T2A and A2T retrieval tasks on benchmark testing sets. Our proposed T-CLAP demonstrates superior performance over the baseline models in most scenarios, particularly across most tasks on the AudioCaps testing set. For tasks performed on the Clotho testing set, the proposed method does not achieve start-of-the-art. This could be attributed to the fact that: (\textit{i}) the majority of the training data consists of 10-second audio clips, while samples with various lengths (e.g., Clotho) only make up a small proportion. This discrepancy may limit the model's ability to generalize across different audio lengths effectively. 
(\textit{ii}) Clotho features longer audio samples that encompass more complex contextual concepts, which may not be captured by the existing retrieval benchmarks. This observation aligns with findings from previous studies~\cite{leverage_audio}, indicating a potential gap in current retrieval metrics to evaluate performance on more contextually rich audio samples.

\begin{table}[htbp]
\caption{The results of zero-shot classification on three benchmark datasets.}
\centering
\small
\resizebox{0.475\textwidth}{!}{%
\begin{tabular}{ccccccc}
\toprule
\multirow{1}{*}{\textbf{Model}}  & \multicolumn{1}{c}{\textbf{ESC-50}} & \multicolumn{1}{c}{\textbf{Urbansound8K}}& \multicolumn{1}{c}{\textbf{VGGSound}} \\
\midrule
Wav2CLIP~\cite{wavclip} &$41.4$ & $40.4$ & $10.0$ \\

AudioClip~\cite{AudioClip} & $69.4$& $65.3$ & - \\

CLAP~\cite{oldclap} & $82.6$ & $73.2$ & - \\

BLAT~\cite{blat} & $80.6$ & $77.3$ & $14.9$ \\


CLAP$_{l}$~\cite{clap} & $91.0$ & $75.8$ & $23.8$ \\


T-CLAP  & $\textbf{96.5}_{\pm0.25}$   & $\textbf{78.4}_{\pm0.38}$ & $\textbf{42.8}_{\pm0.60}$ \\

\bottomrule
\end{tabular}
}
\label{tab:zeroshot}
\end{table}

\noindent \textbf{Zero-shot classification.} To examine the generalization and robustness of the CLAP models, Table~\ref{tab:zeroshot} shows the zero-shot classification results on three benchmark datasets, where T-CLAP outperforms the top five models from previous experiments. Notably, T-CLAP presents a significant gap from the baseline models in the VGGSound~\cite{vggsound} dataset, which consists of larger-scale data with over 300 classes of labels. This significant enhancement could be attributed to the varied and comprehensive content offered by the expansive $2$M training dataset, which improves the robustness of the model.

\begin{table}[htbp]
\caption{The retrieval performance on the temporal feature on the T-Classify task. For Text-to-Audio all three datasets are used (AudioCaps, Clotho, and ESC-mixed-up) while only ESC-mixed-up is applied for Audio-to-Text.}
\centering
\small
\begin{tabular}{ccc}
\toprule

\multirow{2}{*}{\textbf{Model}} & \multicolumn{2}{c}{\textbf{T-Classify}}  \\
  & \multicolumn{1}{c}{\textbf{Text-to-Audio}} & \multicolumn{1}{c}{\textbf{Audio-to-Text}}  \\
\midrule

MMT~\cite{MMt}  & $53.1$ & $55.6$ \\


CLAP$_{m}$~\cite{CLAP-micro}  & $45.7$ & $44.1$ \\

CLAP$_{l}$~\cite{clap}  & $56.2$ & $53.2$ \\

T-CLAP  & $\textbf{87.2}_{\pm0.47}$ & $\textbf{72.0}_{\pm1.35}$  \\
\bottomrule
\end{tabular}
\label{tab:order}
\end{table}

\noindent \textbf{Temproal-featured retrieval.} We use the T-Classify task mentioned in Section~\ref{sec:experiment} to compare the CLAP models on both Text-to-Audio and Audio-to-Text tasks with temporal information. It is noted that only the ESC-mixed-up testing set is applied for Audio-to-Text tasks as this is the only dataset with negative audio clip samples. The results of the baseline models in Table~\ref{tab:order} demonstrate that the previous retrieval models cannot discern the temporal feature, with the overall performance at around 50\%~(selected randomly). Especially in CLAP$_{m}$, the T-classify score does not surpass 50\%, indicating that the model prefers to assign higher similarity scores to negative targets than to correct ones. On the other hand, our proposed approach achieves significant improvements in the T-Classify task. With more than $87\%$ of accuracy on Text-to-Audio tasks, T-CLAP shows an enhanced capability of embedding the audio feature with temporal information. 



\begin{table}[htbp]
\caption{The results of AudioLDM trained with different CLAP as text encoder, where all the models are trained under same configuration and steps. MOS$_{r}$ and MOS$_{t}$ presents human evaluations on overall quality and temporal information accuracy. }
\centering
\small
\resizebox{0.475\textwidth}{!}{%
\begin{tabular}{ccccccc}
\toprule
\multirow{1}{*}{\textbf{Model}}  & \multicolumn{1}{c}{\textbf{KL} $\downarrow$} & \multicolumn{1}{c}{\textbf{FAD} $\downarrow$}& \multicolumn{1}{c}{\textbf{MOS}$_{r} \uparrow$} & \multicolumn{1}{c}{\textbf{MOS}$_{t} \uparrow$}\\
\midrule

AudioLDM$_{M}$ &$2.6_{\pm0.11}$ & $2.4_{\pm0.17}$ & - & -  \\


AudioLDM$_{L}$ & $\textbf{2.2}_{\pm0.13}$ & $2.5_{\pm0.24}$ & $3.5$ & $2.8$  \\

AudioLDM$_{T}$& $2.3_{\pm0.13}$   & $\textbf{1.8}_{\pm0.08}$ & $\textbf{3.7}$ & $\textbf{3.8}$\\

\bottomrule
\end{tabular}
}
\label{tab:audioldm}
\end{table}

\noindent \textbf{Text-to-audio generation.} To compare the effectiveness of temporal-enhanced embedding, we train AudioLDM models using different CLAP models as the condition encoder for $500,000$ steps. AudioLDM$_{M}$, AudioLDM$_{L}$, and AudioLDM$_{T}$ presents the model trained with CLAP-Microsoft~\cite{CLAP-micro}, CLAP-LAION~\cite{clap} and T-CLAP respectively. Table~\ref{tab:audioldm} shows that the AudioLDM model employing T-CLAP achieve better results on text-to-audio benchmarks. Nevertheless, when integrated with T-CLAP encoders, AudioLDM achieves significantly higher scores in the Mean Opinion Score (MOS) for the accuracy of temporal information. Both the objective and subjective evaluations show that the enhanced sequential feature within T-CLAP embedding contributes to the improvement of AudioLDM in terms of both the overall quality and the temporal feature correctness.

\section{Conclusion}
\label{sec:conclusion}
We propose a novel method to enhance the CLAP on temporal information. We first design two pipelines to generate the negative captions and a temporal-focused loss to guide the model. In addition, we introduce a new task, T-Classify, to evaluate the performance of identifying the temporal feature. Compared with the baseline models, T-CLAP provides improved performance in modelling temporal features, as well as achieving state-of-the-art results on retrieval and zero-shot classification tasks. Evaluations on AudioLDM trained with different CLAP models show that the improvements of audio and text embedding in T-CLAP can contribute to performance enhancement in downstream tasks. However, this paper mainly focuses on the situation of sounds, which may result in performance reduction on tasks under other audio domains~(e.g., speech, music). In addition, all the negative audio samples are generated artificially, which affects the robustness of the temporal capability in audio encoders. Future work includes the extension of this method on more audio content, such as speech and music, as well as on more downstream tasks.


\small
\bibliographystyle{IEEEbib}

\begin{thebibliography}{10}

\bibitem{clip}
Alec Radford, Jong~Wook Kim, Chris Hallacy, Aditya Ramesh, Gabriel Goh, Sandhini Agarwal, Girish Sastry, Amanda Askell, Pamela Mishkin, Jack Clark, et~al.,
\newblock ``Learning transferable visual models from natural language supervision,''
\newblock in {\em International Conference on Machine Learning}, 2021.

\bibitem{class1}
Kaiyang Zhou, Jingkang Yang, Chen~Change Loy, and Ziwei Liu,
\newblock ``Conditional prompt learning for vision-language models,''
\newblock in {\em Proceedings of the IEEE/CVF Conference on Computer Vision and Pattern Recognition}, 2022.

\bibitem{class2}
Kaiyang Zhou, Jingkang Yang, Chenchange Loy, and Ziwei Liu,
\newblock ``Learning to prompt for vision-language models,''
\newblock {\em International Journal of Computer Vision}, vol. 130, no. 9, pp. 2337--2348, 2022.

\bibitem{retrieval1}
Aneeshan Sain, Ayan~Kumar Bhunia, Pinaki~Nath Chowdhury, Subhadeep Koley, Tao Xiang, and Yi-Zhe Song,
\newblock ``Clip for all things zero-shot sketch-based image retrieval, fine-grained or not,''
\newblock in {\em Proceedings of the IEEE/CVF Conference on Computer Vision and Pattern Recognition}, 2023.

\bibitem{retrieval2}
Alberto Baldrati, Marco Bertini, Tiberio Uricchio, and Alberto Del~Bimbo,
\newblock ``Effective conditioned and composed image retrieval combining clip-based features,''
\newblock in {\em Proceedings of the IEEE/CVF Conference on Computer Vision and Pattern Recognition}, 2022.

\bibitem{caption1}
Ron Mokady, Amir Hertz, and Amit~H Bermano,
\newblock ``{CLIPcap}: {CLIP} prefix for image captioning,''
\newblock {\em arXiv:2111.09734}, 2021.

\bibitem{caption2}
Yoad Tewel, Yoav Shalev, Idan Schwartz, and Lior Wolf,
\newblock ``Zero-shot image-to-text generation for visual-semantic arithmetic,''
\newblock {\em arXiv:2111.14447}, 2021.

\bibitem{generate1}
Yogesh Balaji, Seungjun Nah, Xun Huang, Arash Vahdat, Jiaming Song, Karsten Kreis, Miika Aittala, Timo Aila, Samuli Laine, Bryan Catanzaro, et~al.,
\newblock ``{EDIFFI}: Text-to-image diffusion models with an ensemble of expert denoisers,''
\newblock {\em arXiv:2211.01324}, 2022.

\bibitem{generate2}
Alexander~Quinn Nichol, Prafulla Dhariwal, Aditya Ramesh, Pranav Shyam, Pamela Mishkin, Bob Mcgrew, Ilya Sutskever, and Mark Chen,
\newblock ``{GLIDE}: Towards photorealistic image generation and editing with text-guided diffusion models,''
\newblock in {\em International Conference on Machine Learning}. PMLR, 2022.

\bibitem{generate3}
Chitwan Saharia, William Chan, Saurabh Saxena, Lala Li, Jay Whang, Emily~L Denton, Kamyar Ghasemipour, Raphael Gontijo~Lopes, Burcu Karagol~Ayan, Tim Salimans, et~al.,
\newblock ``Photorealistic text-to-image diffusion models with deep language understanding,''
\newblock {\em Advances in Neural Information Processing Systems}, vol. 35, pp. 36479--36494, 2022.

\bibitem{clap}
Yusong Wu, Ke~Chen, Tianyu Zhang, Yuchen Hui, Taylor Berg-Kirkpatrick, and Shlomo Dubnov,
\newblock ``Large-scale contrastive language-audio pretraining with feature fusion and keyword-to-caption augmentation,''
\newblock in {\em IEEE International Conference on Acoustics, Speech and Signal Processing}, 2023.

\bibitem{audioldm}
Haohe {Liu}, Zehua {Chen}, Yi~{Yuan}, Xinhao {Mei}, Xubo {Liu}, Danilo {Mandic}, Wenwu {Wang}, and Mark~D. {Plumbley},
\newblock ``{AudioLDM: Text-to-Audio generation with latent diffusion models},''
\newblock in {\em International Conference on Machine Learning}, 2023.

\bibitem{reaudioldm}
Yi~Yuan, Haohe Liu, Xubo Liu, Qiushi Huang, Mark~D Plumbley, and Wenwu Wang,
\newblock ``Retrieval-augmented text-to-audio generation,''
\newblock in {\em International Conference on Acoustics, Speech, and Signal Processing}, 2023.

\bibitem{pre_exp}
Zixian Ma, Jerry Hong, Mustafa~Omer Gul, Mona Gandhi, Irena Gao, and Ranjay Krishna,
\newblock ``{CREPE}: Can vision-language foundation models reason compositionally?,''
\newblock in {\em Proceedings of the IEEE/CVF Conference on Computer Vision and Pattern Recognition}, 2023.

\bibitem{pre_search}
Koustuv Sinha, Robin Jia, Dieuwke Hupkes, Joelle Pineau, Adina Williams, and Douwe Kiela,
\newblock ``Masked language modeling and the distributional hypothesis: Order word matters pre-training for little,''
\newblock in {\em Proceedings of the Conference on Empirical Methods in Natural Language Processing}, 2021.

\bibitem{audiocaps_statis}
Zeyu Xie, Xuenan Xu, Mengyue Wu, and Kai Yu,
\newblock ``{Enhance Temporal Relations in Audio Captioning with Sound Event Detection},''
\newblock in {\em Proc. INTERSPEECH 2023}, 2023, pp. 4179--4183.

\bibitem{htsat}
Ke~Chen, Xingjian Du, Bilei Zhu, Zejun Ma, Taylor Berg-Kirkpatrick, and Shlomo Dubnov,
\newblock ``{HTS-AT}: A hierarchical token-semantic audio transformer for sound classification and detection,''
\newblock in {\em IEEE International Conference on Acoustics, Speech and Signal Processing}, 2022.

\bibitem{roberta}
Yinhan Liu, Myle Ott, Naman Goyal, Jingfei Du, Mandar Joshi, Danqi Chen, Omer Levy, Mike Lewis, Luke Zettlemoyer, and Veselin Stoyanov,
\newblock ``{Roberta}: A robustly optimized bert pretraining approach,''
\newblock in {\em CoRR}, 2019.

\bibitem{oldclap}
Soham Deshmukh, Benjamin Elizalde, and Huaming Wang,
\newblock ``Audio retrieval with {Wavtext5K} and {CLAP} training,''
\newblock in {\em {CoRR}}, 2022.

\bibitem{CLAP-micro}
Benjamin Elizalde, Soham Deshmukh, Mahmoud Al~Ismail, and Huaming Wang,
\newblock ``{CLAP} learning audio concepts from natural language supervision,''
\newblock in {\em IEEE International Conference on Acoustics, Speech and Signal Processing}, 2023.

\bibitem{MMt}
Andreea~Maria Oncescu, A~Koepke, Jo{\~a}o~F Henriques, Zeynep Akata, and Samuel Albanie,
\newblock ``Audio retrieval with natural language queries,''
\newblock in {\em Interspeech}, 2021.

\bibitem{mlase}
Xinhao Mei, Xubo Liu, Jianyuan Sun, Mark~D Plumbley, and Wenwu Wang,
\newblock ``On metric learning for audio-text cross-modal retrieval,''
\newblock in {\em Interspeech}, 2022.

\bibitem{TAP}
Yifei Xin, Dongchao Yang, and Yuexian Zou,
\newblock ``Improving text-audio retrieval by text-aware attention pooling and prior matrix revised loss,''
\newblock in {\em IEEE International Conference on Acoustics, Speech and Signal Processing}, 2023.

\bibitem{countclip}
Roni Paiss, Ariel Ephrat, Omer Tov, Shiran Zada, Inbar Mosseri, Michal Irani, and Tali Dekel,
\newblock ``Teaching {CLIP} to count to ten,''
\newblock {\em arXiv:2302.12066}, 2023.

\bibitem{esc-50}
Karol~J Piczak,
\newblock ``{ESC:} dataset for environmental sound classification,''
\newblock in {\em Proceedings of the ACM International Conference on Multimedia}, 2015.

\bibitem{audiocaps}
Chris~Dongjoo Kim, Byeongchang Kim, Hyunmin Lee, and Gunhee Kim,
\newblock ``{AudioCaps}: Generating captions for audios in the wild,''
\newblock in {\em Annual Conference of the North American Chapter of the Association for Computational Linguistics}, 2019.

\bibitem{clotho}
Konstantinos Drossos, Samuel Lipping, and Tuomas Virtanen,
\newblock ``Clotho: An audio captioning dataset,''
\newblock in {\em IEEE International Conference on Acoustics, Speech and Signal Processing}, 2020.

\bibitem{autoacd}
Luoyi Sun, Xuenan Xu, Mengyue Wu, and Weidi Xie,
\newblock ``A large-scale dataset for audio-language representation learning,''
\newblock {\em arXiv preprint arXiv:2309.11500}, 2023.

\bibitem{ub8k}
J.~Salamon, C.~Jacoby, and J.~P. Bello,
\newblock ``A dataset and taxonomy for urban sound research,''
\newblock in {\em International Conference on Multimedia}, 2014.

\bibitem{vggsound}
Honglie Chen, Weidi Xie, Andrea Vedaldi, and Andrew Zisserman,
\newblock ``{VGGSound}: A large-scale audio-visual dataset,''
\newblock in {\em International Conference on Acoustics, Speech, and Signal Processing}, 2020.

\bibitem{leverage_audio}
Ho~Hsiang Wu, Oriol Nieto, Juan~Pablo Bello, and Justin Salomon,
\newblock ``Audio-text models do not yet leverage natural language,''
\newblock in {\em IEEE International Conference on Acoustics, Speech and Signal Processing}, 2023.

\bibitem{wavclip}
Ho~Hsiang Wu, Prem Seetharaman, Kundan Kumar, and Juan~Pablo Bello,
\newblock ``Wav2clip: Learning robust audio representations from clip,''
\newblock in {\em IEEE International Conference on Acoustics, Speech and Signal Processing}, 2022.

\bibitem{AudioClip}
Andrey Guzhov, Federico Raue, J{\"o}rn Hees, and Andreas Dengel,
\newblock ``{AudioCLIP}: Extending {CLIP} to image, text and audio,''
\newblock in {\em IEEE International Conference on Acoustics, Speech and Signal Processing}, 2022.

\bibitem{blat}
Xuenan Xu, Zhiling Zhang, Zelin Zhou, Pingyue Zhang, Zeyu Xie, Mengyue Wu, and Kenny~Q Zhu,
\newblock ``Blat: Bootstrapping language-audio pre-training based on audioset tag-guided synthetic data,''
\newblock {\em arXiv:2303.07902}, 2023.

\end{thebibliography}

\end{document}